\documentclass[prl,showpacs,preprintnumbers,twocolumn,amsmath,amssymb,superscriptaddress]{revtex4-1}

\usepackage{booktabs}
\usepackage{dsfont}
\usepackage{siunitx}
\usepackage{float}
\usepackage{units}
\usepackage{graphicx}
\usepackage{dcolumn}
\usepackage{bm}
\usepackage{epstopdf}
\usepackage{color}
\usepackage{tabularx}

\begin{document}
\title{Superconducting Energy Scales and Anomalous Dissipative Conductivity in \\Thin Films of Molybdenum Nitride}

\author{Julian~Simmendinger}
\author{Uwe~S.~Pracht}
\author{Lena~Daschke}
\affiliation{1.~Physikalisches Institut, Universit\"{a}t Stuttgart, Pfaffenwaldring 57, 70569 Stuttgart, Germany}

\author{Thomas~Proslier}
\author{Jeffrey~A.~Klug}
\affiliation{Material Science Division, Argonne National Laboratory, Lemont, Illinois 60439, USA}

\author{Martin~Dressel}
\author{Marc~Scheffler}
\email[]{scheffl@pi1.physik.uni-stuttgart.de}
\affiliation{1.~Physikalisches Institut, Universit\"{a}t Stuttgart, Pfaffenwaldring 57, 70569 Stuttgart, Germany}

\date{\today}

\begin{abstract}

We report investigations of molybdenum nitride (MoN) thin films with different thickness and disorder and with superconducting transition temperature \unit[9.89]{K} $\ge{T_c}\ge$ \unit[2.78]{K}. Using terahertz frequency-domain spectroscopy we explore the normal and superconducting charge carrier dynamics for frequencies covering the range from \unit[3 to 38]{cm$^{-1}$} (\unit[0.1 to 1.1]{THz}). The superconducting energy scales, i.e. the critical temperature $T_c$, the pairing energy $\Delta$, and the superfluid stiffness $J$, and the superfluid density $n_s$ can be well described within the Bardeen-Cooper-Schrieffer theory for conventional superconductors. At the same time, we find an anomalously large dissipative conductivity, which cannot be explained by thermally excited quasiparticles, but rather by a temperature-dependent normal-conducting fraction, persisting deep into the superconducting state. Our results on this disordered system constrain the regime, where discernible effects stemming from the disorder-induced superconductor-insulator transition possibly become relevant, to MoN films with a transition temperature lower than at least 2.78\,K.    
\end{abstract}

\maketitle

\section{\label{sec:level1}Introduction}
The fundamental statement of the Anderson theorem \cite{Anderson59} that superconductivity is insensitive to non-magnetic disorder has been contested in recent years: experiments revealed that thin films of strongly disordered superconductors close to the mobility edge exhibit a massively reduced superconducting transition temperature $T_c$ compared to modestly disordered or clean films, and $T_c$ can actually become zero at a critical disorder. Here, the system undergoes a transition from a coherent many-body ground state composed of delocalized Cooper pairs and thermally activated quasiparticle states (a superconductor) to a ground state where the quasiparticle states are incoherent and localized (an insulator) presumably without an intermediate ground state of incoherent but delocalized quasiparticle states (a metal) \cite{Gantmakher2010,Lin2015}. This transition from superconductor to insulator (SIT) has become both a paradigm for a quantum phase transition \cite{Sac11} tuned by a non-thermal control parameter such as disorder or structural granularity and a rich host of intriguing phenomena such as a spatially dependent energy gap \cite{Sacepe2008,Kamlapure2013}, a notable peak in the magnetoresistance \cite{baturina05,baturina07}, scaling \cite{lemarie13} and vortex-charge-duality behavior \cite{ovadia13}, enhanced fluctuations at the resistive transition \cite{liu11,Mondal2011}, and a peculiar gapped density of states above $T_c$ \cite{Sacepe2010,sacepe11,mondal13,chock09}. Owing to their unique properties such as a small electron diffusion coefficient and a low charge carrier density, extremely thin disordered superconductors also play a key role in advanced applications such as superconducting nanowire single photon detectors (SNSPDs) \cite{Matarajan2012,Ilin11,henrich12}.
\begin{figure}[b]
	\centering
	\includegraphics[scale=0.36]{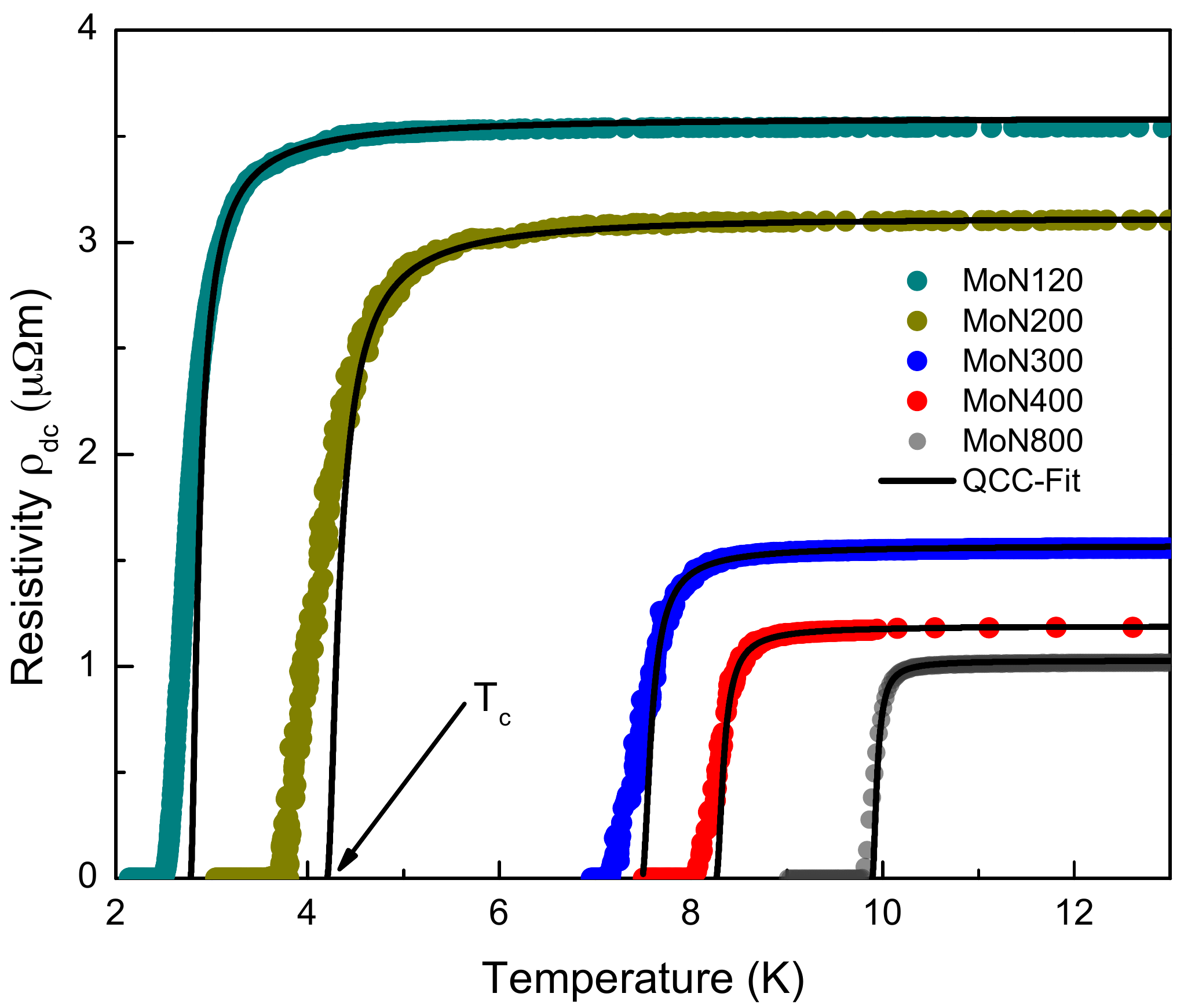}
	\caption{Resistivity $\rho_\text{dc}$ versus temperature for all MoN films under study. With decreasing thickness, the superconducting transition shifts to lower temperatures. The regime of superconducting fluctuations above $T_c$ is well captured by quantum corrections to conductivity (QCC, solid lines), most notably the Aslamazov-Larkin paraconductivity.  Note the resistivity tails below $T_c$ (defined as the temperature where the QCC fit is zero), presumably due to structural inhomogeneity.\label{AlleResistivity}}
\end{figure}
\begin{table*}[t]
\begin{center}
\caption{Overview of the parameters characterizing structural (film thickness $d$, grain diameter $\delta$), free-electron (room-temperature dc-resistivity $\rho_\mathrm{dc}$, Fermi wave-vector $\times$ electron mean free path $k_F\ell$), and superconducting (critical temperature $T_c$, zero-temperature estimates of the energy gap $2\Delta(0)$, superfluid density $n_s(0)$, superfluid stiffness $J(0)$, penetration depth $\lambda(0)$, and the coupling ratio $2\Delta(0)/k_BT_c$) properties of the MoN films under study. The number in the sample name represents the amount of ALD cycles of the films. X-ray reflectivity was used to determine the thickness \cite{Proslier}. \emph{T$_\text{c}$} was obtained from fits of \emph{$\rho_{dc}(T)$} to QCC theory displayed in FIG.~\ref{AlleResistivity}. \label{AllProperties}}
\setlength{\tabcolsep}{2.65mm}
\renewcommand{\arraystretch}{1.5}
\begin{tabular}
{p{1cm}p{0.8cm}p{0.8cm}p{0.8cm}p{0.8cm}p{0.8cm}p{0.8cm}p{0.8cm}p{1.2cm}p{1cm}p{1cm}}
\hline\hline
Name &  \emph{d} (nm) &{$\delta$} (nm) & {$\rho_\mathrm{dc}$ ($\mu\Omega m$)} &{$k_F\ell$}& {\emph{T}$_\emph{c}$ (K)} & 2$\Delta(0)$ (cm$^{-1}$) & $\frac{2\Delta(0)}{\emph{k}_\emph{B}\emph{T}_\emph{c}}$ & {\emph{n$_s$}(0) $\left(10^{25} m^{-3}\right)$} & {\emph{J}(0) (K)} & {$\lambda$(0) ($\mu$m)} \\\hline
MoN120 &  8.4    & 3.1 &  3.20 & 2.02 &  2.78 & 7.21 & 3.73  & 0.76   & 14.0    & 1.93\\
MoN200 &  12.2   & 6.7 &  2.93 & 3.07 &	4.22 & 10.38 & 3.56 & 1.97   & 52.7    & 1.20\\
MoN300 &  15.1   & 9.7 &  1.61 & 6.99 &	7.50 & 16.44 & 3.15  & 5.79   & 192.0   & 0.70\\
MoN400 &  17.4   & 11.6 &  1.34 & 10.1 &	8.26 & 20.28 & 3.52  & 3.52  & 134.5 & 0.90\\
MoN800 &  29.6   & 15.7 &  1.26 & 13.04 &	9.89 & 23.24 & 3.41  & 9.57  & 621.9  & 0.54\\\hline\hline
\end{tabular}  
\end{center}
\end{table*}

One of the most intriguing open and fundamental questions concerns the mechanism of the SIT: Is it the loss of Cooper pairs or the loss of the coherent superfluid that suppresses $T_c$ towards the SIT? The answer to this question is hidden in the disorder-evolution of the superconducting energy scales, i.e. the pairing amplitude $\Delta$ and the superfluid stiffness $J$, which is a measure for the robustness of the superfluid phase-coherence against fluctuations. The lesser of both scales determines $T_c$. In order to address this question, we employ optical measurements in the THz frequency range which have turned out to be a powerful approach to access $\Delta$ and $J$ of thin-film superconductors \cite{Pracht2016} as just one asset of optical spectroscopy in this context \cite{Steinberg2008,Degiorgi2009,Xi2010,Driessen2012,Scheffler2015}. Our material of choice for the present study are thin films of disordered molybdenum nitride (MoN), where the SIT has not yet been observed, the suppression of $T_c$ with decreasing thickness $d$ and increasing electrical resistivity $\rho_{dc}$, however, resembles well-established SIT systems such as NbN, InO, or TiN.

\section{\label{sec:level2}Samples and Experiment}
Bulk MoN is a conventional superconductor with a $T_c$ up to 12\,K \cite{Marchand1999,Bull2004}. In this work, we study several thin polycrystalline films of different thickness $d$ ranging from 8.4 to 29.6\,nm with a roughness of less than 3\,nm \cite{Groll2014,Klug2013,Proslier}. The average grain size $\delta$ increases from 3.1 to 15.7\,nm going from the thinnest (8.4\,nm) to the thickest (29.6\,nm) film. The crystal sizes were extracted from XRD measurements using the Scherrer formula that relates the diffraction peak width to the crystallite size. The films are grown by atomic layer deposition (ALD) on $5\times5$\,mm$^2$ (001)-silicon substrates covered with a native oxide layer. The residue chlorine concentration (from the MoCl$_5$ precursor used) is negligible and varies between 0.2 and 0.5\% as found from Rutherford back-scattering and x-ray photo-electron measurements, respectively, and does not depend on $d$. With decreasing $d$ we observe an increase of dc-transport resistivity $\rho_{dc}$ and a concomitant reduction of $T_c$ from 9.89 to 2.78\,K, see Table~\ref{AllProperties}. With decreasing thickness we found a reduction of the product of Fermi wave-vector and electron mean free-path $k_F\ell$ quantifying the effective degree of disorder for films grown on quartz substrates, which are from the structural and electronic point of view identical with films grown on silicon as shown by X-ray diffraction and transport measurements. Consequently, one is free to use decreasing values of $T_c$ and $d$, or increasing values for  $\rho_{dc}$ as descriptive measures of growing disorder. We note, however, that this selective assignment is not assured \emph{a priori} and may not apply for films beyond this work. Throughout this paper, we refer to the samples as MoN\emph{x}, where \emph{x} is the number of ALD cycles, which governs the film thickness. We measured the dc-transport resistivity $\rho_{dc}$ in four-point geometry, and extract $T_c$ as fit parameter within the theory of quantum corrections to conductivity (QCC), capturing the superconducting fluctuations as it has been established previously \cite{Sacepe2010,Baturina2012}. The resistive transitions of all samples measured are plotted in Fig.~\ref{AlleResistivity}. We observe a broadening of the transition with decreasing thickness and $T_c$, which suggests  an increase of superconducting fluctuations as expected for increasing disorder. Fluctuations of the Aslamazov-Larkin type are most prominent, while the other corrections \cite{Sacepe2010,Baturina2012} play a minor role. A closer examination, however, reveals resistive tails, which cannot be accounted for by QCC fluctuations, as shown in Fig.~\ref{AlleResistivity}, but may result from structural inhomogeneity \cite{Benfatto2009}, as discussed below.
Apart from transport measurements, we performed optical spectroscopy to measure the complex transmission for frequencies 3 to 38\,{cm$^{-1}$ (\unit[0.1 to 1.1]{THz}) utilizing a Mach-Zehnder interferometer
equipped with backward-wave oscillators as tunable sources of continuous and  monochromatic THz radiation and a Golay-cell or $^4$He-bolometer as detectors. With a home-built optical $^4$He-cryostat we performed measurements down to 1.7\,K \cite{Pracht2016,Hering2007}. For more experimental details, see Ref. \cite{Kozlov1998,Dressel2008,Pracht2013}.


\section{\label{sec:citeref}Methods}

Typical spectra of transmission amplitude $t$ and phase shift normalized to frequency $\phi/\nu$ (\emph{relative} phase shift), where $\nu$ = $\omega/2\pi$ is the frequency, of a MoN sample measured in the normal and superconducting states are shown in Fig.~\ref{MoN300Tr+Ph}(a) and (b). The pronounced oscillation pattern stems from multiple reflections inside the substrate, which acts as Fabry-P\'{e}rot resonator \cite{Pracht2013}. While this pattern is constant in the normal state, it changes drastically below $T_c$, which calls for a strong frequency dependence of the optical properties of the film.
To model the particular behavior of $t$ and $\phi/\nu$, we use the Fresnel equations for multiple reflections, where the thickness $d$ and dielectric function $\hat{\epsilon}(\nu)=\epsilon_1(\nu)+i\epsilon_2(\nu)$ of substrate (s) and thin film (f) directly enter
\begin{eqnarray}\label{FresnelEq1}
t&=&t(d_\mathrm{s},\epsilon_1^\mathrm{s}(\nu),\epsilon_2^\mathrm{s}(\nu);\,d_\mathrm{f},\epsilon_1^\mathrm{f}(\nu),\epsilon_2^\mathrm{f}(\nu))\\
 \phi/\nu&=&\phi(d_\mathrm{s},\epsilon_1^\mathrm{s}(\nu),\epsilon_2^\mathrm{s}(\nu);\,d_\mathrm{f},\epsilon_1^\mathrm{f}(\nu),\epsilon_2^\mathrm{f}(\nu))/\nu
 \label{FresnelEq2}
\end{eqnarray} 
Equivalently, this can be expressed in terms of the dynamical conductivity $\hat{\sigma}(\nu)=\sigma_1(\nu)+i\sigma_2(\nu)$, which is directly related to $\hat{\epsilon}$ via
\begin{equation}\label{permittivityEq}
\hat{\epsilon}(\nu)=1+\frac{i}{2\pi}\frac{\hat{\sigma}(\nu)}{\nu\epsilon_0}
\end{equation} 
with $\epsilon_0$ the permittivity of the vacuum.\\
To disentangle the properties of substrate and film, a bare substrate was measured beforehand and its optical parameters were determined to be $\epsilon_1^\mathrm{s}=11.7$ and $\epsilon_2^\mathrm{s}=0$ independent of frequency and temperature in the range studied in this work. In what follows, we will neglect the superscripts in $\sigma^\mathrm{f,s}_{1,2}$ and always refer to the conductivity of the MoN film.

\section{Results}
\subsection{Anomalous dissipative conductivity}
\begin{figure}[t]	
	\centering
	\includegraphics[scale=0.45]{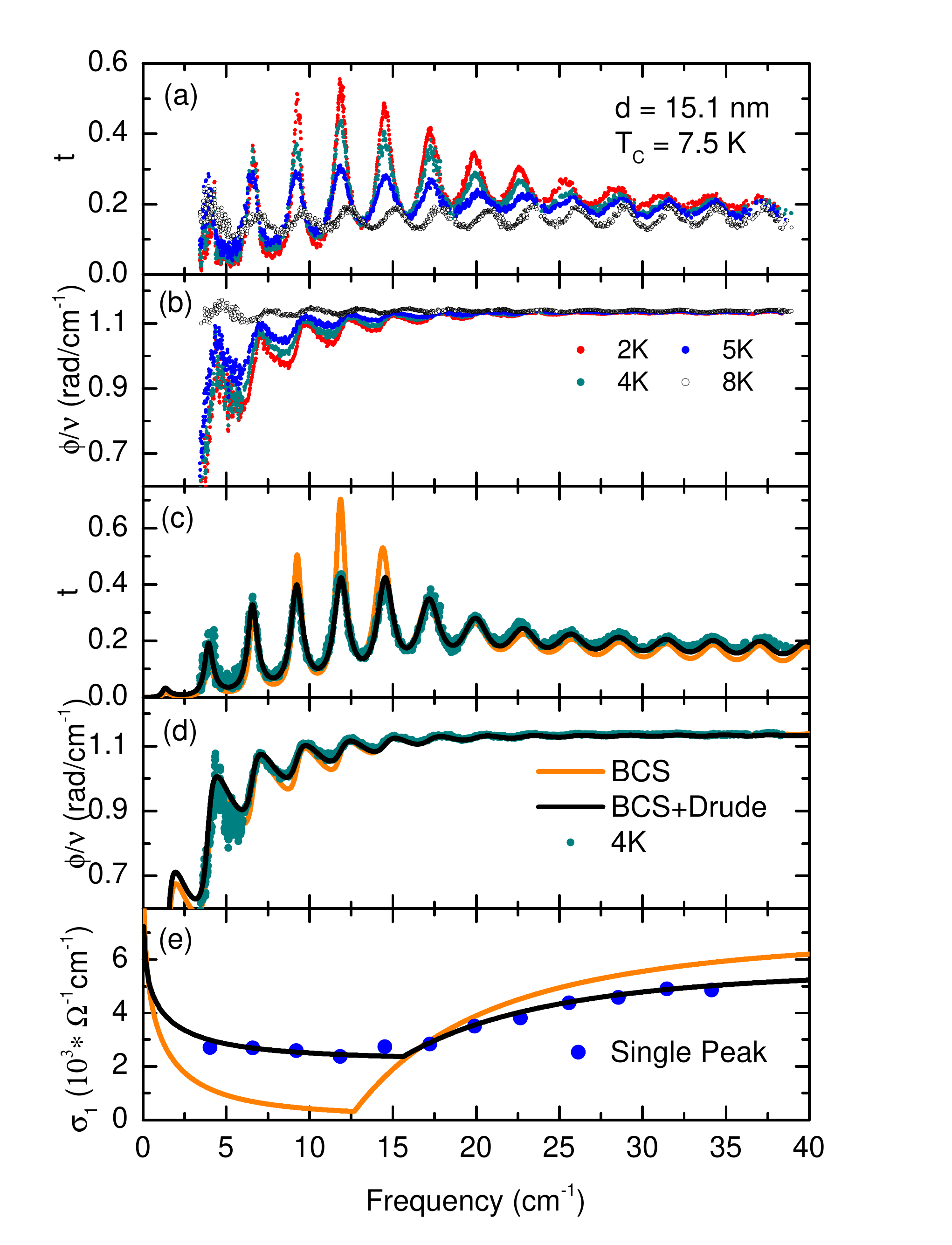}
	\caption{\label{MoN300Tr+Ph}(a) Raw transmission amplitude and (b) relative phaseshift of a representative sample (MoN300) versus frequency at various temperatures in the superconducting state and above $T_c=7.5$\,K. The strong increase of $t$ below $T_c$ is due to the opening of the SC gap and reduction of free electrons with dissipative dynamics. Panels (c) and (d) display $t$ and $\phi/\nu$ together with fits to MB Eqs.~(\ref{s1}) and (\ref{s2}) solely (orange curve) and to MB supplemented with an additional Drude term Eqs.~(\ref{eq:sigma1}) and (\ref{eq:sigma2}) (black curve). Clearly, only with the additional dissipative contribution a proper fit is possible. Panel (e) displays the real part $\sigma_1(\nu)$ of the dynamical conductivity at $T=4$\,K in the SC state calculated from $t$ and $\phi/\nu$ with the corresponding fits to MB and MB+Drude.   }
\end{figure}
We first focus on the normal state at a temperature slightly above $T_c$ and the regime of superconducting fluctuations. Here, we find $t$ and $\phi/\nu$ of all samples to be well described by $\sigma_1(\nu)$ and $\sigma_2(\nu)$ of a normal metal, i.e. by the Drude (D) formula \cite{Dressel2002}
\begin{eqnarray}
\sigma_{1}(\nu)=\sigma_{1}^\mathrm{D}(\nu)&=&\frac{\sigma_\mathrm{dc}}{1+(2 \pi\nu\tau)^2}\label{eq:sigma1}\\
\sigma_{2}(\nu)=\sigma_{2}^\mathrm{D}(\nu)&=&\frac{\sigma_\mathrm{dc} \, 2 \pi\nu\tau}{1+(2 \pi\nu \tau)^2}\label{eq:sigma2}
\end{eqnarray} 
where $\sigma_\mathrm{dc}$ is the dc-transport conductivity, $\tau$ the scattering time and $\gamma=1/\tau$ the scattering rate. Fig.~\ref{MoN300Tr+Ph} displays the raw $t$ and $\phi/\nu$ of a representative sample with $T_c=7.5$\,K. Apart from the pronounced Fabry-P\'{e}rot oscillations, both $t$ and $\phi/\nu$ at \unit[8]{K} are dispersionless, which is in agreement with a scattering rate $\gamma$ at frequencies much higher than the studied spectral range. The same result is found for all samples and the corresponding values of $\rho_\mathrm{dc}=\sigma^{-1}_\mathrm{dc}$ are shown in Table~\ref{AllProperties}.\\
For a conventional BCS superconductor below $T_c$ in the dirty limit, $\hat{\sigma}(\nu)$ follows the Mattis-Bardeen (MB) equations \cite{Mattis1958}, which describe the dynamics of both the superfluid condensate and the thermally excited quasiparticles
\begin{eqnarray}
\label{s1}
\frac{\sigma_1^\mathrm{MB}(\nu)}{\sigma_n}&=&\frac{\pi e^2n_s}{m^*\sigma_n}\delta(\nu)+\frac{2}{h\nu} \int\limits_\Delta^\infty \mathrm{d}\epsilon\, g(\epsilon)\left[ f_\epsilon- f_{\epsilon-h\nu}\right]\nonumber\\
&&-\frac{\Theta(h\nu-2\Delta)}{h\nu}\int\limits_{\Delta-h\nu}^{-\Delta}  \mathrm{d}\epsilon\, g(\epsilon)) \left[1-2 f_{\epsilon+h\nu}\right]\\
\label{s2}
\frac{\sigma_2^\mathrm{MB}(\nu)}{\sigma_n}&=&\frac{1}{h\nu} \int\limits_{-\Delta,\Delta-h\nu}^\Delta  \mathrm{d}\epsilon\,\Big(g(\epsilon)\left[ 1-2f_{\epsilon+h\nu}\right]\nonumber\\
&&\times\frac{\epsilon(\epsilon+h\nu)+\Delta^2}{\sqrt{\Delta^2-\epsilon^2}\sqrt{(\epsilon+h\nu)^2-\Delta^2}}\Big),
\end{eqnarray}

\noindent where the function $g(\epsilon)$ and $f_\epsilon$ are
\begin{align}
\label{ge}
g(\epsilon)&=\frac{\epsilon(\epsilon+h\nu)+\Delta^2}{\sqrt{\epsilon^2-\Delta^2}\sqrt{(\epsilon+h\nu)^2-\Delta^2}}\\
f_{\epsilon}&=\frac{1}{\mathrm{exp}\left(\frac{\epsilon-\mu}{k_BT}\right)+1}
\end{align}     
and $\Delta$ is the superconducting energy gap, $m^*$ and $e$ are the effective carrier mass and charge, $\Theta$($\epsilon$) is the Heaviside step function, $\mu$ is the total chemical potential, $n_s$ the superfluid density, and $\sigma_n$ the dc-transport conductivity right above $T_c$. Usually, any finite dissipative conductivity in the SC state is attributed to unpaired quasiparticles and captured by the second term in Eq.\,(\ref{s1}), so that $\sigma_n=\sigma_\mathrm{dc}$  using the notation from above. For the MoN thin films under study, however, a model based on Eq.\,(\ref{s1})  turns out to be insufficient. If we do not restrict ourselves to weak-coupling SC, the only free parameter in Eq.~(\ref{s1}) is $\Delta$, whose variation alone does not lead to a reasonable fit, see the orange curve in Fig.~\ref{MoN300Tr+Ph}. The strongest deviation appears at frequencies around $2\Delta$, where the actual transmission amplitude is considerably smaller than the fit. This implies that in addition to the thermally excited quasiparticles another dissipative channel is present. While various complex mechanisms leading to finite in-gap absorption in disordered SC have been recently addressed in both theory \cite{Swa14,cea2014,cea15} and experiment \cite{crane2007,Bac14,sherman15}, here, we can model  $t$ and $\phi/\nu$  reasonably well by simply adding a Drude contribution to the MB dynamics
\begin{equation}
\sigma_{1,2}(\nu)=\sigma_{1,2}^\mathrm{MB}(\nu)+\sigma_{1,2}^\mathrm{D}(\nu)
\label{MB+D}
\end{equation}    
where the normal-state values of $\tau$ and $\sigma_\mathrm{dc}$ in Eqs.~(\ref{eq:sigma1}) and (\ref{eq:sigma2}) are replaced by $\tilde{\tau}$ and $\tilde{\sigma}_\mathrm{dc}$. We note that for a vanishing dc-transport resistance it is sufficient to have a single percolative superconducting path bypassing normal-conducting ones, whereas an optical measurement is still sensitive to non-superconducting areas because here the dissipation integrated over the entire volume is probed. Fig.~\ref{MoN300Tr+Ph}(c) and (d) exhibits $t$ and $\phi/\nu$ exemplary at $T=4$\,K in the superconducting state together with a fit to Fresnel equations with charge carrier dynamics described by Eq.~(\ref{MB+D}). In this particular case, a (frequency-independent) Drude behavior with $\tilde{\sigma}_\mathrm{dc}=1.5\times10^{-3}$\,$\mu\Omega$cm and $\tilde{\tau}\to 0$ (meaning that the scattering rate $\tilde{\Gamma}$=1/$\tilde{\tau}$ is considerable higher than the spectral range of the present study, and $\sigma_2^{\text{D}}$ is negligibly small) fits $t$ and $\phi/\nu$ very well.
\begin{figure}
\begin{centering}
\includegraphics[scale=0.35]{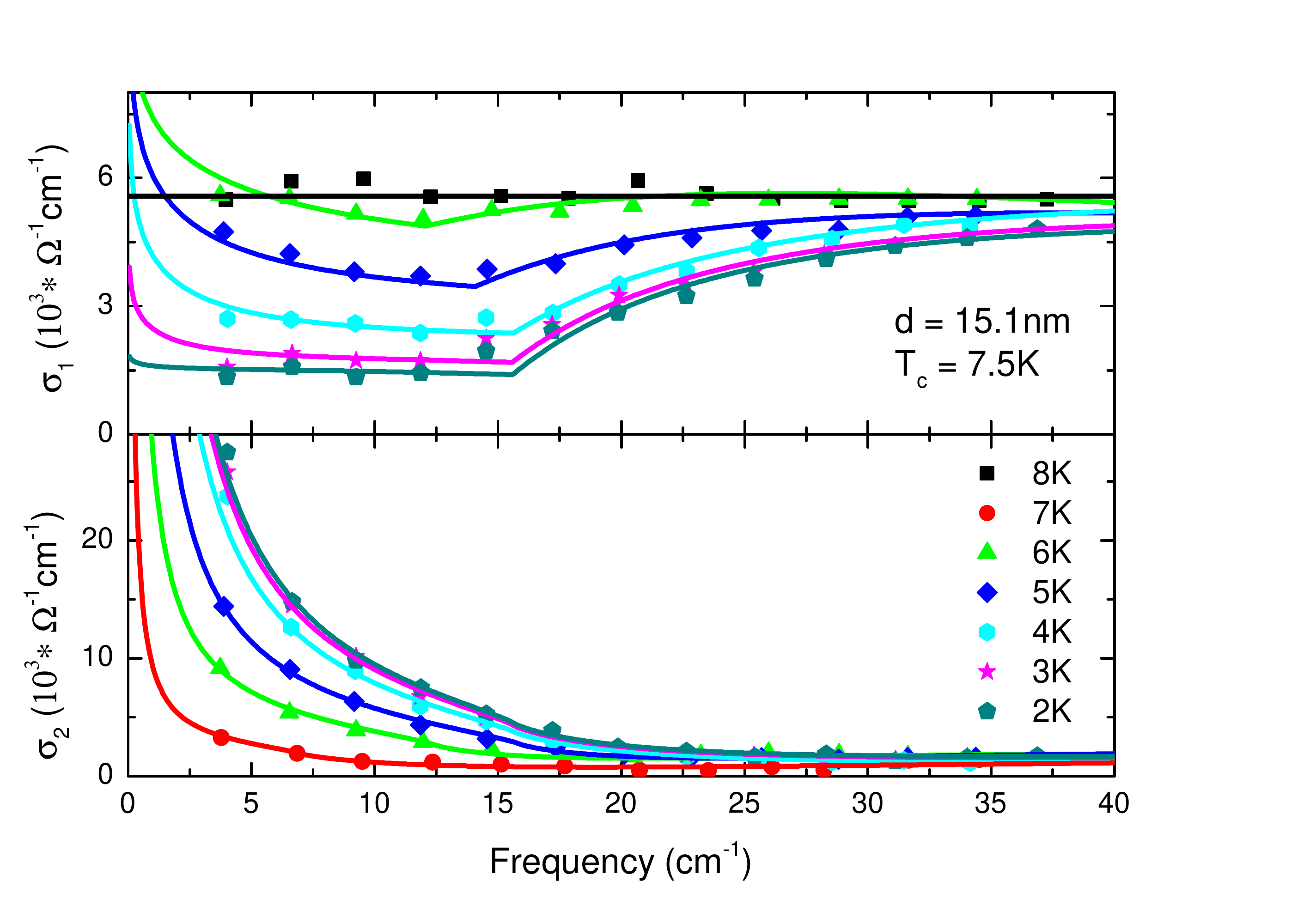}
\caption{\label{S1+S2} Real part (upper panel) and imaginary parts of the dynamical conductivity of a representative sample (MoN300) in the normal state and at various temperatures below\\ $T_c=7.5$\,K. The solid lines are fits to Eq.~(\ref{s1}) and (\ref{s2}).}
\end{centering}
\end{figure} 
The conductivity offset is more obvious when looking directly at $\sigma_1(\nu)$ of the film rather than $t$ and $\phi/\nu$. To obtain  $\sigma_1(\nu)$ and $\sigma_2(\nu)$, we employ a \emph{single-peak analysis} and fit $t$ and $\phi/\nu$ in a narrow window around each Fabry-P\'{e}rot resonance by the Fresnel equations \cite{Pracht2013}. For each Fabry-P\'{e}rot resonance located at a frequency $\nu_i$ we obtain a pair or $\epsilon_{1,2}(\nu_i)$ or, equivalently, $\sigma_{1,2}(\nu_i)$ with a frequency dependence that is not restricted to any particular microscopic model such as Eqs.~(\ref{eq:sigma1})-(\ref{s2}). Fig.~\ref{S1+S2} displays $\sigma_1(\nu)$ and $\sigma_2(\nu)$ of the same sample shown in Fig.~\ref{MoN300Tr+Ph} obtained  from the single-peak analysis for various temperatures $T<T_c$. The additional Drude contribution is obvious in the dissipative conductivity, where, in this particular case, the measured $\sigma_1(\nu)$ does not fall below $1.5 \times 10^3$~$\mu\Omega$cm even at our lowest temperature of 2~K. This is in disagreement with a conventional superconductor, where at $T/T_c\approx 0.25$ one would expect almost no dissipation inside the spectral gap.\\ 
The behavior of charge carriers in a conventional superconductor is sketched in FIG.~\ref{SketchBCSvsDrude+BCS}(a), where only Cooper-pairs and thermally excited quasiparticles are present below $T_c$. However, in our MoN films, we find charge carriers with metallic properties even below $T_c$. Thus we suggest a model for superconducting MoN featuring metallic states below $T_c$ as it is sketched in FIG.~\ref{SketchBCSvsDrude+BCS}(b).
\begin{figure}[H]	
	\centering
	\includegraphics[scale=0.30]{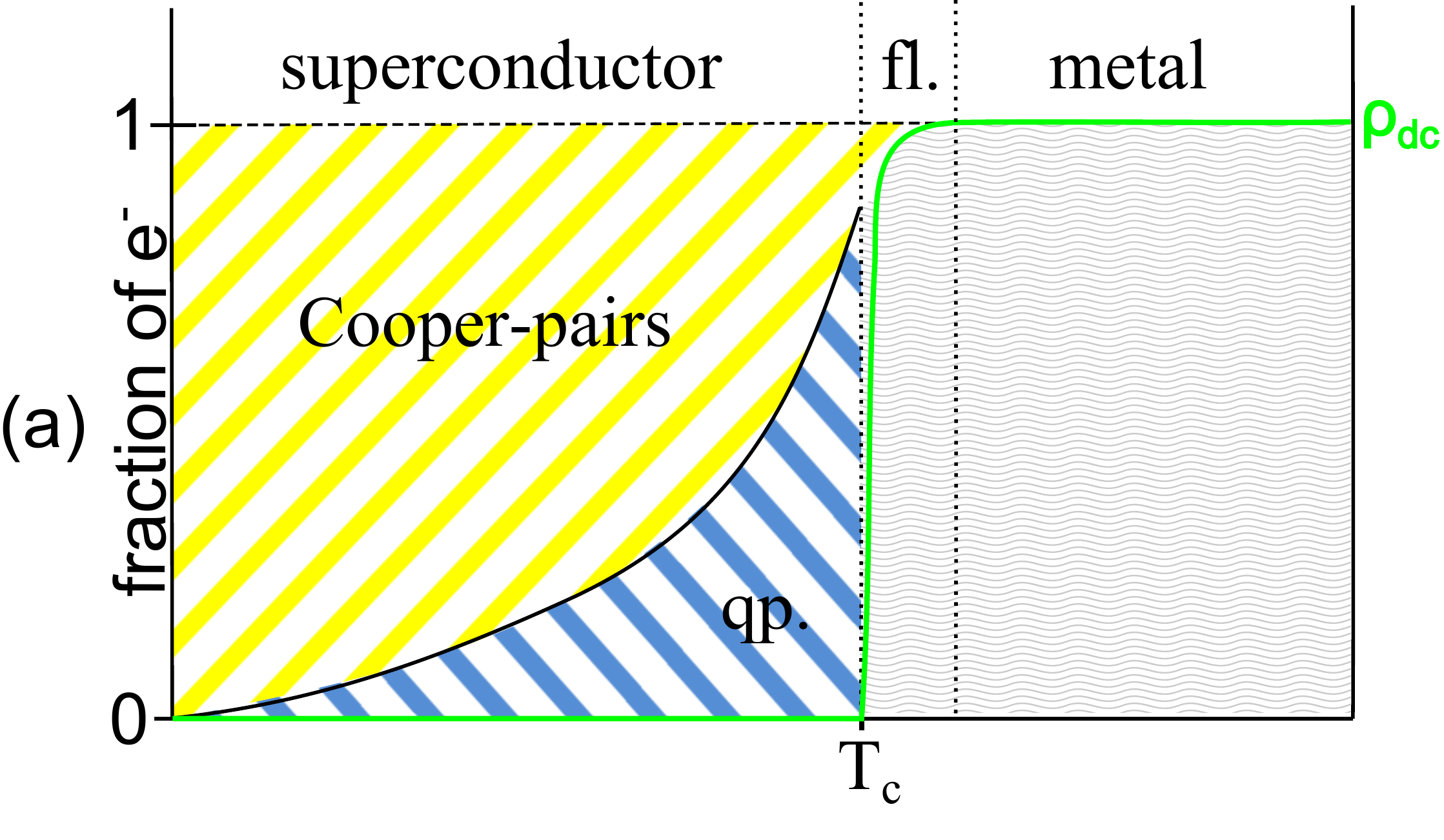}
	\quad
	\includegraphics[scale=0.30]{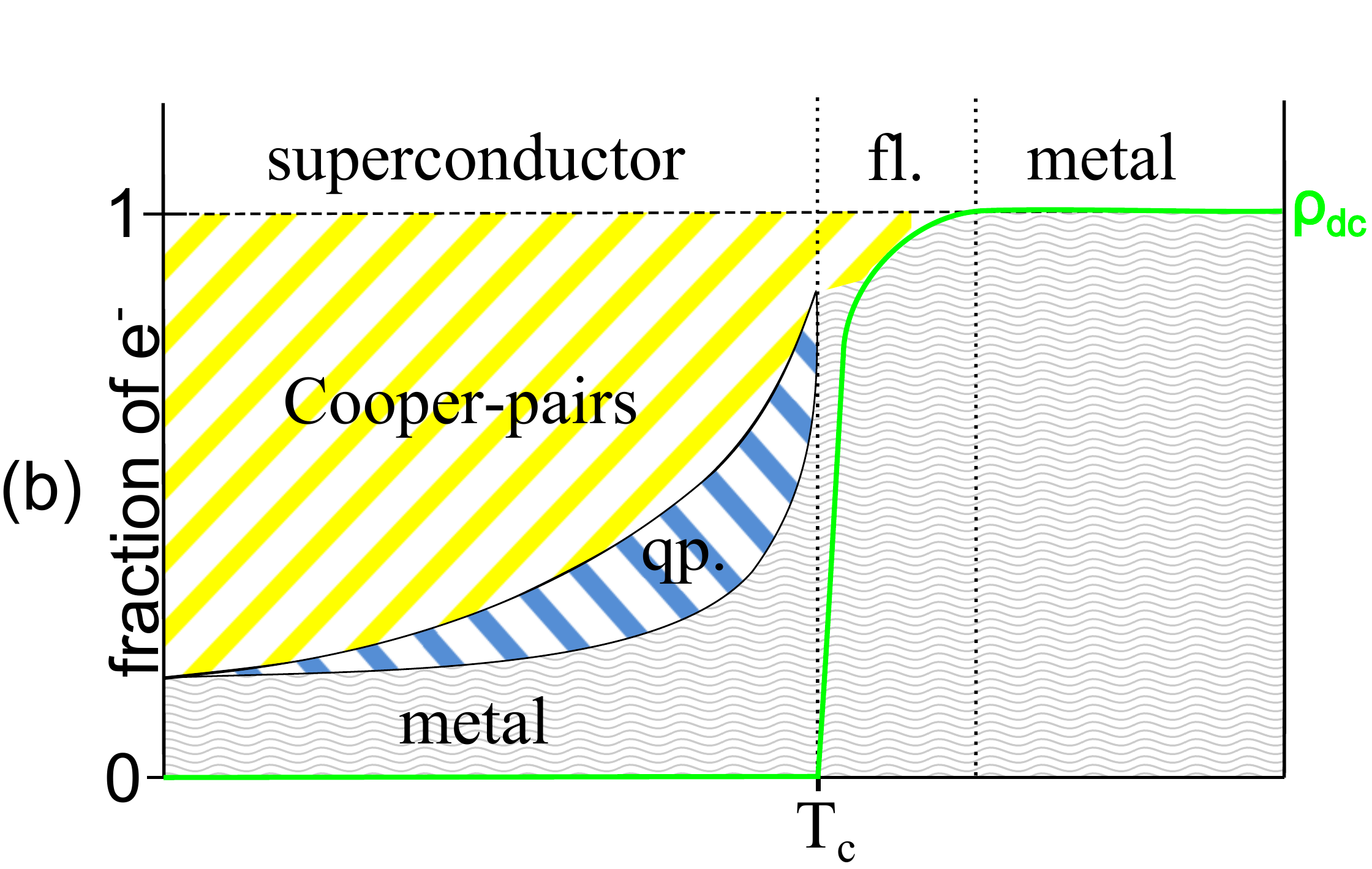}
	\caption{Total amount of charge carriers distributed into Cooper-pairs, thermally excited quasiparticles (qp.), and non-condensing metallic electrons (e$^-$), schematically shown for (a) a normal BCS material and (b) a model for MoN, where normal conducting (metallic) electrons are present in the globally superconducting state at all temperatures down to $T=0$. Above $T_c$, in the fluctuation regime (fl.) there is a small amount of Cooper-pairs present. The temperature-dependent
dc resistivity is also shown schematically \label{SketchBCSvsDrude+BCS}}
\end{figure}
It is important to stress that this phenomenon is not restricted to a single film, but turns out to be a general feature of all MoN films studied, and furthermore, the temperature-dependence of $\sigma^\mathrm{D}$ can be fitted by a phenomenological function
\begin{equation}\label{exponential}
\frac{\sigma^{\text{D}}(T)}{\sigma_\mathrm{dc}} = a~\mathrm{exp}\left\{b\frac{T}{T_c}\right\}
\end{equation}
Fig.~\ref{DrudeContr} displays $\sigma^\mathrm{D}(T)$ of all samples studied in this work together with fits by Eq.~(\ref{exponential}). Values of the fit parameters $a$ and $b$ we listed in Table~\ref{ExponentialFit}. Neither $a$ nor $b$ render a discernible thickness dependence. 
\begin{table}[t]
\begin{center}
\caption{Parameters $a,b$ obtained by fitting $\sigma^\mathrm{D}/\sigma_\mathrm{dc}$ by Eq.~(\ref{exponential}).\label{ExponentialFit}}
\setlength{\tabcolsep}{0.9mm}
\renewcommand{\arraystretch}{1.4}
\begin{tabular}
{cS[table-format=4.2]S[table-format=4.2]}\hline\hline
sample & $a$  & $b$ \\\hline
MoN120  &  0.22  & 1.36  \\
MoN200 &  0.14 & 1.67  \\
MoN300 & 0.16  & 1.76  \\
MoN400 &  0.21 & 1.33  \\
MoN800 &  0.10  & 2.21  \\\hline\hline
\end{tabular}
\end{center}
\end{table}
One conceivable explanation attributes the dissipative contribution to intrinsic inhomogeneity. The assigned $T_c$ is the mean-field temperature, where superconducting percolation across the sample sets in, while separate regions may become superconducting already at a higher critical temperature. Similarly, some regions may remain normal-conducting down to temperatures well below $T_c$. As temperature drops, more and more regions eventually become superconducting, and the remaining dissipative contribution shrinks. Indeed, thin films of TiN and NbN with only marginal disorder and no structural inhomogeneity do not show a finite $\sigma^\mathrm{D}$ contribution in the superconducting state \cite{Pracht2012,Pracht2013}.    
Note, that this inhomogeneity does not necessarily require structural inhomogeneities, but may result from homogeneous disorder leading to an electronically inhomogeneous state as it was shown  directly \cite{Sacepe2008,Kamlapure2013} and indirectly \cite{Sacepe2010,Mondal2011,mondal13} for similar thin films of TiN and NbN. In addition \cite{Benfatto2009}, an inhomogeneous superfluid density, as it may result from either structural or electronic inhomogeneity, causes anomalous tails in the temperature dependence of the dc resistivity that themselves cannot be attributed to superconducting or Berezinskii-Kosterlitz--Thouless type fluctuations. As shown in Fig.~\ref{AlleResistivity}, we indeed observe such resistive tails for all samples, which cannot be captured by fluctuations. Given that this tail is present irrespective of thickness calls for an intrinsic structural inhomogeneity rather than emergent electronic inhomogeneity usually relevant only at strong disorder near the superconductor-insulator transition \cite{Sacepe2008,Kamlapure2013,Sacepe2010,Mondal2011,mondal13}. The ALD growth of MoN thin films generally initiates with the growth of a \unit[1-2]{nm} layer of Mo$_2$N before MoN growth sets in. The enhancement of absolute dissipative conductivity $\sigma_{\text{dc}}(T\to0)$ with increasing MoN film thickness suggests that the ubiquitous \unit[1-2]{nm} Mo$_2$N layer does not serve as explanation for the anomalous dissipation but may only contribute a small universal offset.
\begin{figure}[t]
\begin{centering}
\includegraphics[scale=0.72]{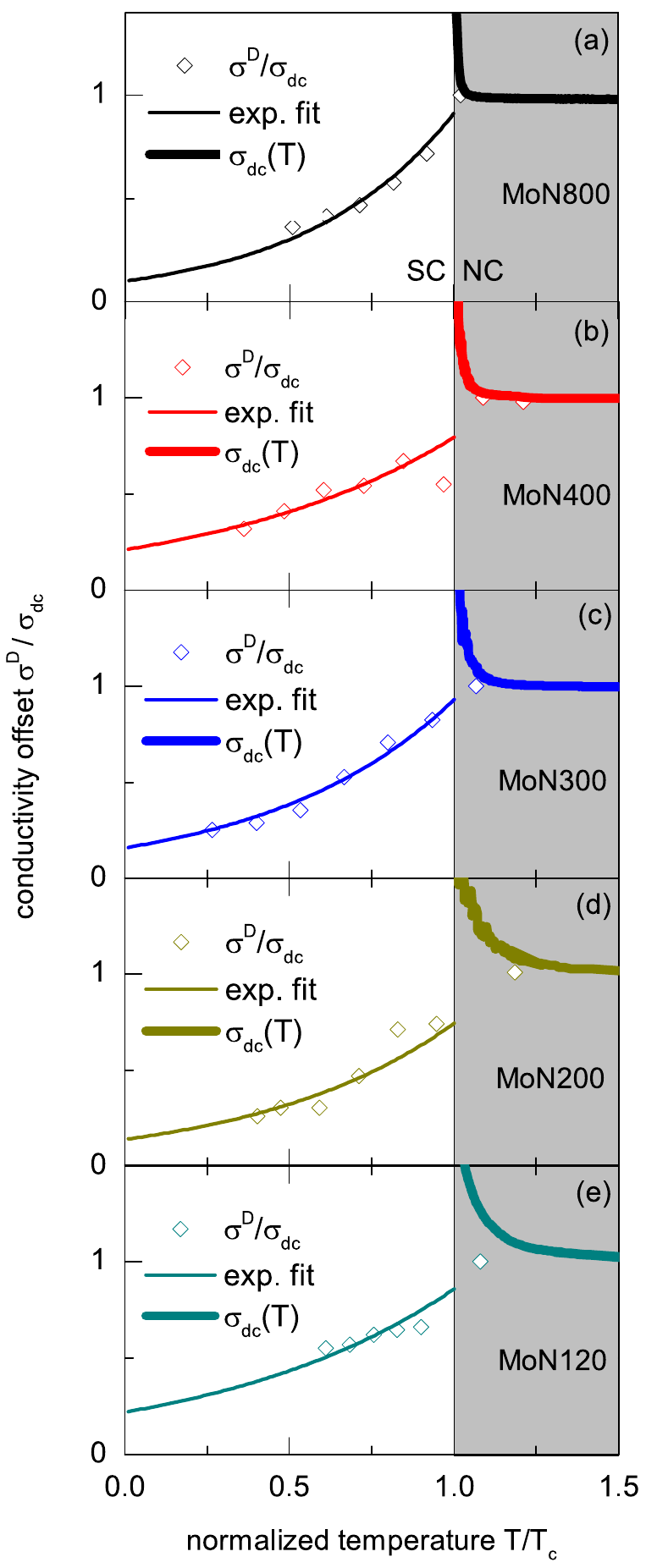}
\caption{\label{DrudeContr}Temperature dependence of the dissipative conductivity normalized to the normal-state conductivity, $\sigma^\mathrm{D}/\sigma_\mathrm{dc}$,  inside the globally superconducting state (white background). The dc-conductivity $\sigma_\mathrm{dc}(T)=\rho_\mathrm{dc}^{-1}$ measured in transport is shown in each panel as thick lines, while the open symbols are obtained from fits of the the optical measurements according to Eq.~(\ref{MB+D}). The thin solid lines are fits according to the phenomenological function (\ref{exponential}).  
}
\end{centering}
\end{figure}     

\begin{figure}[b]
\centering
\includegraphics[scale=0.5]{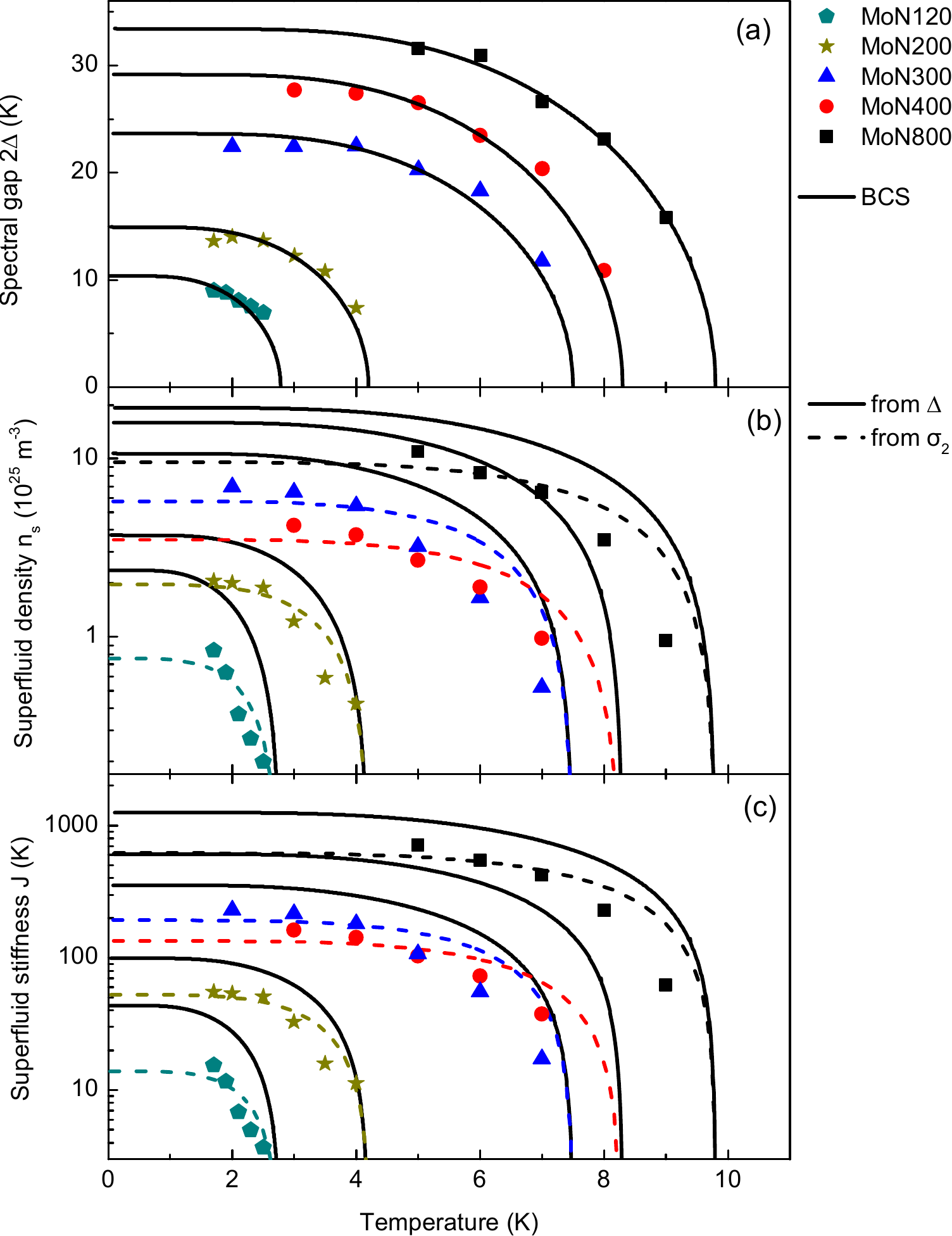}
\caption{\label{D,n,l,J}Temperature dependence of the (a) spectral gap $2\Delta$, (b) superfluid density $n_s$, and (c) stiffness $J$ of all MoN samples under study. Solid lines are fits within BCS theory as described in the text. The dashed lines are calculations of $n_s(T)$ based on the $\Delta(T)$ fits overestimating the actual $n_s$ due to the dissipative conductivity $\sigma^\mathrm{D}$.}
\end{figure}
\subsection{Pairing amplitude, superfluid density, and \\superfluid stiffness}
We now turn to the superconducting contribution in Eq.~(\ref{MB+D}), and the energy scales, i.e. pairing energy $\Delta$ and superfluid stiffness $J$, and the superfluid density $n_s$, which we extract from $\hat{\sigma}(\nu)$ within Mattis-Bardeen theory. We obtain $\Delta$ by fitting $\sigma_1(\nu)$ by Eq.~(\ref{s1}) and (\ref{MB+D}), where $\Delta$ enters as threshold energy of pair-breaking, the spectral gap $2\Delta$. In Fig.~\ref{D,n,l,J}(a) we plot the temperature evolution of $2\Delta$ for all samples together with a fit to the universal BCS behavior of $2\Delta$ obtained by solving the self-consistency equation\cite{Likharev1979}
\begin{equation}
\label{usadel}
\ln\frac{T_c}{T}=2\pi k_B T \sum_{\omega_n}\left[\frac{1}{\hbar \omega_n}-\frac{1}{\sqrt{(\hbar \omega_n)^2+\Delta^2}}\right]
\end{equation}
where $\omega_n=\pi k_B T(2n+1)$ with $n\in \mathds{N}$ are the Matsubara frequencies. The good agreement between theory and experiment allows to extrapolate $2\Delta(T=0)$ and calculate the ratio $2\Delta(0)/k_BT_c$ , see Table~\ref{AllProperties}, which is close to the weak-coupling prediction of 3.53 for all samples. 
The second energy scale, $J$, is closely related to $n_s$, which, on the one hand, withdraws from direct access being the weight of the $\delta(0)$ function in the superfluid response of  Eq.~(\ref{s1}). On the other hand, the superfluid condensate dominates [$\sigma_1(\nu\approx 0) \approx \delta(\nu)]$ and therefore, following Kramers-Kronig relations, also determines $\sigma_2$ at small, but finite frequencies. Considering the Kramers-Kronig transform for the superfluid contribution to $\hat\sigma$
\begin{equation}
\sigma_2(\omega\approx0)=-\frac{2}{\pi}\mathrm{P}\int\limits_0^\infty\frac{\mathrm{d}\omega^\prime\omega}{\omega^{\prime 2}-\omega^2}\frac{\pi n_s e^2\delta(\omega^\prime)}{2m^*}
\end{equation} 
one finds the relation
\begin{equation}
n_s=\frac{2\pi m^*}{e^2}\nu\sigma_2(\nu)\Bigr|_{\nu = 0} \label{ns}
\end{equation}
using $\omega=2\pi\nu$ and $\delta(\omega)=\frac{1}{2\pi}\delta(\nu)$. Fig.~\ref{D,n,l,J}(b) displays $n_s$ of all samples versus temperature obtained from Eq.~(\ref{ns}). The dashed lines are fits to the two-fluid approximation\cite{Dressel2002}, and also here the good agreement between theory and experiment allows to reliably extrapolate $n_s(0)$ at zero temperature. Starting with the thinnest film, $n_s$ rises as thickness increases. With the BCS temperature-dependence of $\sigma_2(T)$,
\begin{equation}
\sigma_2(T,\nu)=\frac{\pi \Delta(T)}{\rho_\mathrm{dc}h\nu}\tanh\left\{\frac{\Delta(T)}{2k_BT}\right\}
\end{equation}
inserted into Eq.~(\ref{ns}) one readily obtains
\begin{equation}
n_s(T)=\frac{2\pi^2 m}{\rho_\mathrm{dc}e^2h}\Delta(T)\tanh\left\{\frac{\Delta(T)}{2k_BT}\right\}\label{nSBCS}
\end{equation}
which explains the rise of $n_s$ with increasing thickness by the simultaneous increase of $\Delta$ and decrease of $\rho_\mathrm{dc}$. A calculation of $n_s(T)$ via Eq.~(\ref{nSBCS}) and $\Delta(0)$ as obtained from $\sigma_1(\nu)$, leads to the solid lines in Fig.~\ref{D,n,l,J}(b) and an overestimation of the superfluid density. This can be understood as natural consequence of the dissipative offset: While a constant value of $\sigma^\mathrm{D}$ below $T_c$ does not change the location of the pair-breaking absorption threshold in $\sigma_1(\nu)$ and consequently leaves $\Delta$ unaffected, the actual superfluid density is reduced by the number of metallic electrons not participating in the superfluid state. Therefore, considering the preservation of the spectral weight, the value of $n_s$ obtained directly from $\sigma_2$ should be smaller than the one calculated from $\Delta$.  \\
Fig.~\ref{D,n,l,J}(c) displays the superfluid stiffness as calculated from $n_s$ via \cite{Benfatto2009}
\begin{equation}
J(T) =  \frac{\hbar^2 n_{s}(T) d}{4m} = 0.62 \times \frac{d [\text{\AA}]}{\lambda^2[(\mu\text{m})^2]}\, \text{K}.
\end{equation}
with $\lambda\propto1/\sqrt{n_s}$ the superconducting penetration depth and $d$ the film thickness. While the temperature dependence of $n_s$ and $J$ is identical up to a numerical factor, the latter quantity has the dimension of energy and can be compared to the other relevant energy scales. The good agreement of $\Delta(T), n_s(T)$, and $J(T)$ with the BCS theory and extensions thereof down to our most-disordered sample constrains the regime where effects stemming from the SIT become relevant to samples with even lower $T_c<2.78$\,K. 
Fig.~\ref{MasterplotJ,Delta,Tc,Ratio} displays the zero-temperature extrapolations of the superconducting energy scales, $J(0),~\Delta(0)$, and $T_c$ as function of resistivity $\rho_{dc}$ (film thickness). For better comparison, all scales are expressed in units of temperature.  
 \begin{figure}[ht]
	\centering
	\includegraphics[scale=0.5]{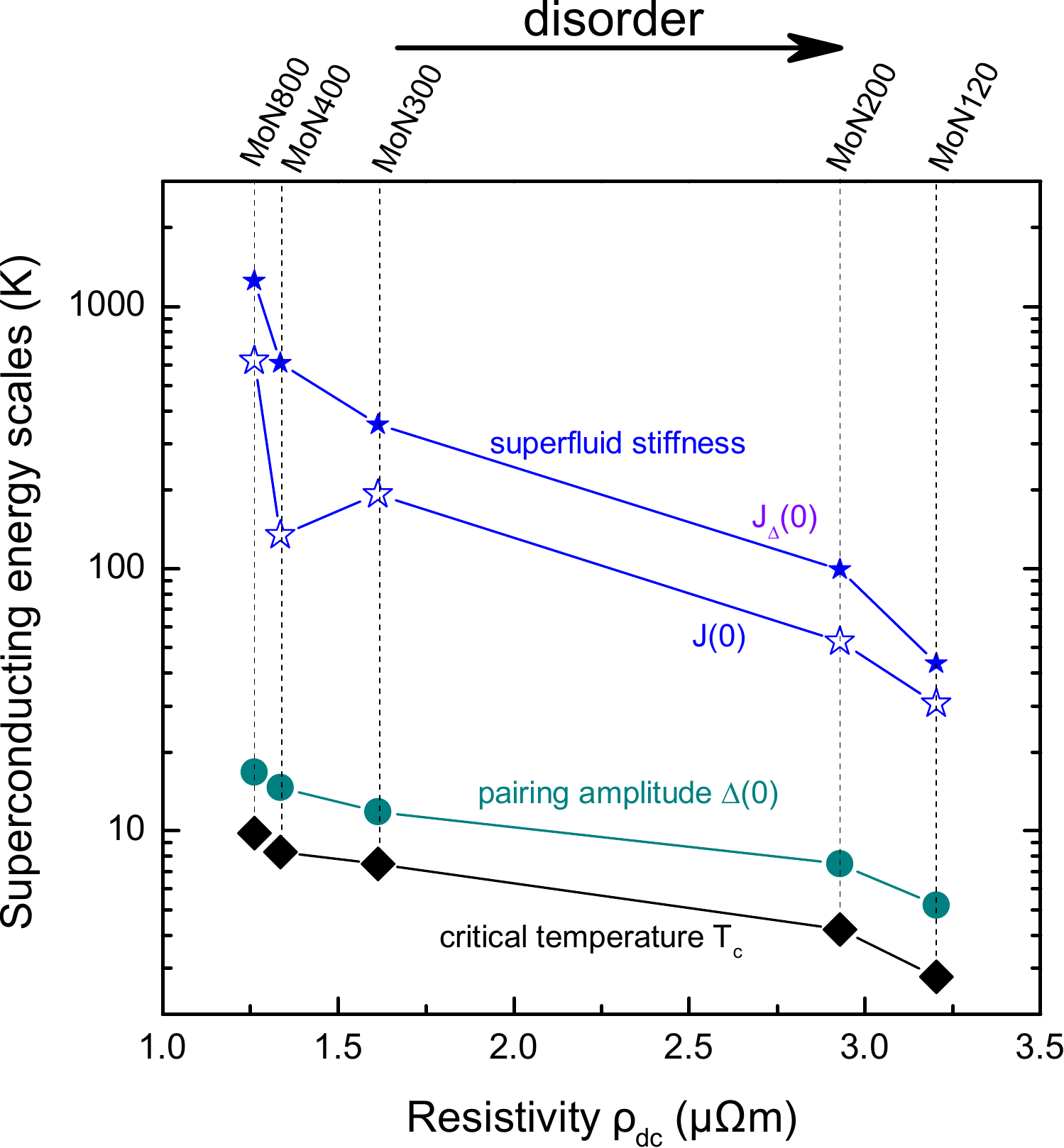}
	\caption{Superconducting energy scales $T_c$, $\Delta(0)$, and $J(0)$ extrapolated to zero-temperature as described in the text versus resistivity $\rho_\mathrm{dc}$ for all MoN samples under study. $T_c$ and $\Delta(0)$ decrease in the same fashion with increasing  $\rho_\mathrm{dc}$ giving a nearly constant weak-coupling ratio, see Table~\ref{AllProperties}. $J(0)$ decreases by nearly two orders of magnitude in the examined  $\rho_\mathrm{dc}$ range but does not fall below $\Delta(0)$, signaling amplitude-driven superconductivity for all our MoN thin-films. \label{MasterplotJ,Delta,Tc,Ratio}}
\end{figure}
With increasing $\rho_{dc}$, all energy scales show a clear decrease, which is strongest for $J(0)$. This behavior is in agreement with the Mattis-Bardeen theory, where $J(0)$ is related to $\Delta(0)$ via
\begin{equation}
J_{\Delta}(0) = \frac{\hbar d}{e^2 \rho_{dc}} \frac{\pi \Delta(0)}{4}
\end{equation}
where the additional factor $d/\rho_{dc}$ causes the stronger reduction. Nevertheless, we find $\Delta(0)<J(0)$ for all films in agreement with an amplitude-driven transition, where superconductivity ceases when the pairing amplitude $\Delta$ becomes zero at $T_c$, i.e. due to the loss of pairing rather than phase-coherence.
The mismatch between $J(0)$ as calculated from $\sigma_2(\nu)$ and $J_\Delta(0)$, see  Fig.~\ref{MasterplotJ,Delta,Tc,Ratio}, is a direct consequence of the dissipative conductivity $\sigma^\mathrm{D}$.   

\section{\label{sec:citeref}Conclusion}
We studied the dynamical conductivity $\hat{\sigma}(\nu)=\sigma_1(\nu)+i\sigma_2(\nu)$ of a series of disordered MoN thin films with different thickness (and resistivity $\rho_{dc}$) ranging from 8.4 to 29.6~nm by means of THz spectroscopy and electrical transport, and extracted the pairing amplitude $\Delta$, the superfluid density $n_s$, and the superfluid stiffness $J$ within BCS theory. The temperature dependence of $\Delta, n_s$, and $J$ is well described within the conventional theory and the evolution with $\rho_{dc}$ suggests superconductivity to cease at $T_c$ due to the loss of pairing rather than phase coherence. The real part $\sigma_1(\nu)$ shows an anomalous dissipative contribution $\sigma^\mathrm{D}$ in the superconducting state that cannot be accounted for by ordinary quasiparticle dynamics within the Mattis-Bardeen (MB) theory. This frequency-independent contribution is found for all MoN samples under study and is suppressed with decreasing temperature. In addition, the superfluid density $n_s$ calculated from $\sigma_2(\nu)$ is smaller than predicted by MB theory for corresponding pairing amplitude. Together with tails in the resistivity curves below $T_c$, all these findings suggest the presence of normal-conducting regions surviving into the globally superconducting state, possibly due to structural inhomogeneity, as explanation of the anomalous optical properties. The agreement of the superconducting properties with the BCS and MB theory holds for all samples, which restricts effects stemming from the SIT to samples with $T_c$ lower than at least 2.78\,K. Our characterization of superconducting MoN thin films is of interest for the development of applications such as microwave resonators \cite{Goeppl2008,Scheffler2013,Singh2014,Ohya2014} or ultra-sensitive photon detection \cite{Matarajan2012,Gao2012,Szypryt2015} and it can serve as reference for studies on other superconductors where the role of inhomogeneity is presently discussed \cite{Gantmakher2010,Lin2015,Stewart07,Eley12,Han14,Biscaras13}. At the same time, further studies using local probes, e.g. scanning tunneling microscopy, are highly desirable to test the idea of inhomogeneous superconductivity on MoN thin films.  

We acknowledge discussion with Lara Benfatto, Ina Schneider, and Christoph Strunk. U.S.P. thanks the Studienstiftung des Deutschen Volkes for financial support.
J.A.K. and T.P. acknowledge support from the Department of Energy, Office of Sciences, Office of High Energy Physics, Early Career Award FWP 50335.

%

\end{document}